\documentclass[12pt]{article}

\usepackage{amsmath}
\usepackage{amssymb}

\def\rdots{\mathinner{\mkern1mu\raise1pt\vbox{\kern1pt\hbox{.}}\mkern2mu
   \raise4pt\hbox{.}\mkern2mu\raise7pt\hbox{.}\mkern1mu}}
\newcommand{\Z}{{\rm Z\kern-.35em Z}}
\newcommand{\bP}{{\rm I\kern-.15em P}}
\newcommand{\Q}{\kern.3em\rule{.07em}{.65em}\kern-.3em{\rm Q}}
\newcommand{\R}{{\rm I\kern-.15em R}}
\newcommand{\h}{{\rm I\kern-.15em H}}
\newcommand{\C}{\kern.3em\rule{.07em}{.65em}\kern-.3em{\rm C}}
\newcommand{\T}{{\rm T\kern-.35em T}}

\newcommand{\be}{\begin{equation}}
\newcommand{\ee}{\end{equation}}

\newcommand{\la}{\lambda}

\newcommand{\pa}{\partial}

\newcommand{\ra}{\rightarrow}

\newcommand{\La}{\Lambda}

\newcommand{\al}{\alpha}
\newcommand{\nn}{\nonumber}

\begin{document}
%\font\twelverm=cmr12
%\font\cs=CMSSI12

\openup 1.5\jot
\centerline{Hidden Structure in Tilings, Conjectured Asymptotic Expansion}
\centerline{for $\la_d$ in Multidimensional Dimer Problem}
\bigskip

\bigskip

\vspace{1in}
\centerline{Paul Federbush}
\centerline{Department of Mathematics}
\centerline{University of Michigan}
\centerline{Ann Arbor, MI 48109-1043}
\centerline{(pfed@umich.edu)}

\vspace{1in}

\centerline{\underline{Abstract}}

\ \ \ \ \ The dimer problem arose in a thermodynamic study of diatomic molecules, and was abstracted into one of the most basic and natural problems in both statistical mechanics and combinatoric mathematics.  Given a rectangular lattice of volume $V$ in $d$ dimensions, the dimer problem loosely speaking is to count the number of different ways dimers (dominoes) may be laid down on the lattice to completely cover it.  It is known that the number of such coverings is roughly $e^{\la_d V}$ for some number $\la_d$.  The first terms in the expansion of $\la_d$ have been known for about thirty years
\[	\la_d \sim \frac 1 2 \; ln(2d) - \frac 1 2 \ .   \]
Herein we present a mathematical argument for an asymptotic expansion
\[	\la_d \sim \frac 1 2 \; ln(2d) - \frac 1 2 + \frac 1 8 \frac 1 d + \frac5{96} \; \frac1{d^2}  + \ \cdots \ . \]
with the first few terms given explicitly.

In a previous paper, [1], we worked with tiles for which there was a weighting function assigned to the tiles.  The weighting function satisfied a normalization condition.  We then associated a quantity we called the ``pressure" to the tilings of a lattice with such tiles.  Loosely speaking, the result of the paper was that if we varied the weighting function, letting it get smoother and smoother (that is, more slowly varying), the corresponding pressures approached a limit.  The limit was, of course, the pressure associated to the ``infinitely smooth weighting function" a constant function.  Herein we continue the statistical mechanics approach one step further.  We construct for any weighting function the {\it perturbation expansion} for the partition function that arises naturally from treating the difference between the actual weighting function and the infinitely smooth weighting function as a perturbation.

We construct a second ``partition function" associated to the first partition function, which we develop in a ``cluster expansion".   Kernels $J_1,J_2,....$ arise in a standard way from the cluster expansion development.  We next specialize to the dimer problem.  We relate our two partition functions and we herein give a formal derivation of an expansion we hope in the future to prove asymptotic.
\[	\la_d \sim \ \frac 1 2 \ ln(2d) - \frac 1 2 + \frac{c_1}d + \frac{c_2}{d^2} + \cdots \  . \]
We know in fact  $c_1 = 1/8$ and $c_2= 5/96$. The $c_i$ are computable from the $J's$.  The leading term $\frac 1 2 \; ln(2d) - \frac 1 2$ does indeed arise as the zeroth order term in the cluster expansion.

The present work was inspired by, and is a natural sequel to, our previous paper [1].  However it is essentially independent thereof.  We write the paper so it may be read without reference to [1], some notation is modified, but we strive to keep consistency between the two papers.

We work with a $d$-dimensional lattice, $\La$, on a torus.  We let $N$ be the number of vertices of $\La$, (hopefully without confusion, we also let $\La$ be the set of vertices).  We consider tiles, $t$, of size, or area, $n$.  (Our tiles need not be connected, and $N/n$ must be an integer.)  We have a weighting function, $f$, on tiles that is normalized so that
\be	\sum_t \; f(t) = \frac 1 n    \ee
where the sum is over all tiles.  A {\it located tile}, ${\it s}_\al$, is a tile placed at a particular location on the lattice, and is exactly a subset of the lattice of cardinality $n$,
\be	{\it s}_\al = \left\{ x^\al_1, ..., x^\al_n\right\} .   \ee
( A tile may be viewed as an equivalence class of subsets under translation, and placing a tile at a particular location corresponds to selecting an element of the equivalence class.)  We denote the corresponding tile as $\bar {\it s}_\al$.  Equation (1) can be written as
\be
\begin{array}[t]{c}
{\displaystyle\sum} \\
{\scriptstyle{{\it s}_\al} } \\
{\scriptstyle {x_0 \in {\it s}_\al} }
\end{array}
f\big(\bar {\it s}_\al \big) = 1
\ee
The sum in (3) is over subsets of $\La$ of cardinality $n$ that contain a fixed point of the lattice $x_0$.  This equation is easily identified with equation (3) of [1]; as naturally for any located tile $s_\al$ we identify
\be	
f\big( \bar {\it s}_\al \big) = f\big( x^\al_1, ..., x^\al_n \big)
\ee
where the left side $f$ is of this paper, and the right side $f$ is as in [1].

We now consider a tiling, $T_i$, of $\La$ by tiles of size $n$.  $T_i$ may be identified with a set of $N/n$ located tiles
\be	T_i = \left\{ s^i_1, s^i_2, ..., s^i_{N/n} \right \}  \ee
Since $T_i$ is a tiling
\be	\bigcup_\al \; s^i_\al = \La	\ee
and
\be	s^i_\al \cap s^i_\beta = \phi, \ \ \al \not= \beta \ .   \ee
The partition function, $Z$, and the quantity we call the pressure, $p$, are given by the equation
\be	Z = e^{Np} = \sum_{T_i} \prod_{s_\al \in T_i} f(\bar s_\al) .   \ee
The right side is the sum over tilings, each tiling weighted by the product of the weight functions of the tiles appearing in the tiling.

The ``infinitely smooth weight function" is constant on tiles.  We let $f_0$ be its value on any tile, easily calculated to be
\be	f_0 = \frac{(n-1)!(N-n)!}{(N-1)!} \ .		\ee
We write
\begin{eqnarray}
f(\bar s_\al) &=& f_0 + \Big( f(\bar s_\al) - f_0 \big) \nn \\
  \\
&=& f_0 + v(\bar s_\al) \nn
\end{eqnarray}
where we will treat $v$ as a perturbation.  Equation (8) is now
\be	Z = e^{Np} = \sum_{T_i} \prod_{s_\al \in T_i} \Big( f_0 + v(\bar s_\al) \Big).   \ee
We expand $Z$ in terms of powers of $v$
\be	Z= Z_0 + Z_1 + Z_2 + . . . 	\ee
$Z_i$ homogeneous of degree $i$ in the $v's$.
Singling out $Z_0$ first
\be	Z_0 = \sum_{T_i} \prod_{s_\al \in T_i} f_0 = \sum_{T_i} f^{N/n}_0	\ee
and one calculates
\be	\sum_{T_i} = \frac{N!}{(\frac N n)!(n!)^{N/n}} .  \ee
We put together (9), (13), and (14) to define
\be	Z_0 = e^{N\hat p^0(N)}   \ee
where
\be	\lim_{N\ra \infty} \hat p^0(N) = \frac{1-n}n.	\ee
(These are equations (8) and (9) of [1].)

We turn for a moment to the dimer problem.[2]   \ We relate the quantity $\la_d$ of the dimer problem to our variables (if our $f$ is set equal a constant on dimers and zero on other tiles).
\be		e^{N\la_d} = (2d)^{N/2} \; Z	\ee
or
\be		\la_ d = \frac 1 2 \; ln(2d) + \frac 1 N \ ln \; Z	\ee
the $(2d)^{N/2}$ in equation (17) arises since our normalization condition (1) requires our $f$ to be $\frac 1{2d}$ on each dimer, whereas to just count tilings (with no weighting) would correspond to $f=1$ on each dimer.  Replacing $Z$ by $Z_0$ in (18) and taking the limit $N \ra \infty$ one gets
\be   \la_d \cong \frac 1 2 \ ln(2d) - \frac 1 2 \ .  \ee
Our expressions for $\la_d$ will all be in the infinite volume limit.  Thus taking our zeroth order approximation for $Z, Z=Z_0$, we get the approximation for $\la_d$ of (19), the result of [2]!  In [2] there are bounds on the error of this approximation.

Returning to the general case we factor out $Z_0$ from $Z$ in (12)
\begin{eqnarray}
Z &=& Z_0 Z^*  \\
Z^* &=& 1 + Z^*_1 + Z^*_2 + ..... \\
Z^*_i &=& Z_i \big/ Z_0
\end{eqnarray}
There is a natural factoring of $Z_i$ into a contribution from the factors of $v$ in (11) which we call $\bar Z^*_i$ and the factors of $f_0$ in (11) which we call $\beta(N,i) Z_0$ so that
\be	Z_i = \beta(N,i)Z_0\bar Z^*_i	\ee
and thus
\be   Z^*_i = \beta(N,i) \bar Z^*_i	\ee
$\bar Z^*_i$ is defined below, and it is a tedious calculation to show the surprising result that $\beta(N,i) \ra 1$ as $N \ra \infty$.  We let $\tilde Z^*$ be $Z^*$ with $\beta(N,i)$ replaced by 1.
\be	\tilde Z^* = 1 + \bar Z^*_1 + \bar Z^*_2 + . . . 	\ee
We turn to the detailed specification of $\bar Z^*_i$
\be
\bar Z^*_i = \frac 1{i!} \begin{array}[t]{c}
{\displaystyle\sum} \\
{\scriptstyle{s_1,s_2,...,s_i }} \\
{\scriptstyle {{\rm disjoint }}}
\end{array}
\prod^i_{\al = 1} v(\bar s_\al)
\ee
We are trying to put our expression for $\tilde Z^*$ into a form that can be identified with equation (2.5a) from [3].  Some knowledge of the structure of cluster expansions becomes necessary.  Article [3] is a standard reference, but a knowledgeable patient friend may be more helpful.  

With the notation in [3], the formal cluster expansion for $\tilde Z^*$ is given as a translation for equation (2.7) of [3].
\be	ln \ \tilde Z^* = \sum_s \frac 1{s!} \ J_s	\ee
\be	J_s = \sum_{s_1,s_2,...,s_s} \ v(\bar s_1) ... v(\bar s_s) \psi'_c(s_1,s_2,...,s_s)	\ee
The located tiles appearing in the sum for $J_s$ are forced to overlap so that they cannot be divided into two disjoint sets of located tiles.  $\psi'_c$ is a numerical factor depending on the pattern of overlaps.  An illustrative computation of a sample $J_s$ is treated in the appendix.  

The $J_s$ for the dimer problem easily satisfy
\be   J_1 = 0 \ee
and we will show in a succeeding paper that $J_s$ is of the form
\be		J_s = \frac{C_r}{d^r} + \frac{C_{r+1}}{d^{r+1}} + \cdots + \frac{C_s} {d^{(s-1)}} \  .  	\ee
with $r \ge s/2$.

We want the asymptotic behavior of $Z^*$ as $N \ra \infty$
\be	Z^* = \sum_i \beta(N,i) \bar Z^*_i  \ee
and we know from (27) the asymptotic behavior as $N \ra \infty$ of $\tilde Z^*$ 
\be	\tilde Z^* = \sum_i  \bar Z^*_i  \ee
We argue the asymptotic behavior of (31) arises from a largest term of the form
\be   \beta\big(N, \Sigma\; i\; \al_i \; N\big) \bar J^{\al_1N}_1 \cdots \bar J^{\al_{s+1}N}_{s+1} \cdot 
\frac{N^{\Sigma \; \al_i N}}{(\al_1N)! \cdots (\al_{s+1} N)!}  \ee
{\it provided} all these $J's$ are $\ge 0$.  Here $\bar J_i N = (1/i!)J_i$.  We are finding that portion of the asymptotic behavior due to just $J_1, ..., J_{s+1}$.  We choose the $\al_i$ to maximize (33).  We will use (via another tedious calculation)
\[   \beta(N, jN) \sim e^{N\big[ \big(\frac{1-2j}{2}\big) ln(1-2j)+j\big]}  \]

Differentiating (33) with respect to the $\al_i$ leads to the equations
\be	ln \; \al_k = ln \; \bar J_k + \frac \pa{\pa \al_k} \left[ \left( \frac{1-2 \Sigma i \al_i}{2} \right) ln (1 - 2\Sigma \; i \al_i) + 
\Sigma \; i \al_i  \right]  \ee
or
\be	\al_k = \bar J_k e^{F_k(\al's)}   \ee
where (34) and (35) define the $F_k$.  Equation (35) is solved for $\al_k$ as a formal power series in the $\bar J$ by iterating starting from setting the $\al's = 0$ on the right side of equation.  Again, in a succeeding paper, we argue these same formulas hold even if some of the $J's$ are negative.

We substitute (35) into (33) to get
\be
Z^* \ \sim \ e^{N \Big\{ - \Sigma \;\al_i\; F_i \;+\; \Sigma \; \bar J_i \; e^{F_i}\;+\; \frac{1-2\Sigma \;i \;\al_i}{2} \; ln \big(1-2\Sigma \; i \;\al_i\big) 
+ \Sigma i \al_i  \Big\} } .
\ee
From (34), (35), and (31) we can expand the exponent in (36) into a formal power series in the $\bar J_i$, and get the asymptotic series
\be   \la_d \sim \ \frac 1 2 \ ln(2d) - \frac 1 2 + \frac{c_1}d + \frac{c_2}{d^2} + \cdots  \ee
and from
\be		\bar J_2 = \frac 1 8 \; \frac 1 d, \bar J_3 = \frac 1{12} \; \frac 1{d^2}, \ \bar J_4 = - \; \frac 3{64} \ \frac {(2d-1)}{d^3}\;,	\ee
get
\be		c_1 = \frac 1 8, \; c_2 = \frac 5{96} \  .   \ee

$\la_2$ was calculated exactly in [4] and [5].  Good bounds for $\la_3$ are given in [6].  We hope the asymptotic expansion we have developed actually is convergent, even for $d=2$ and $d=3$!  We are actively working to show this.

\noindent
\underline{Note}.  In an earlier form of this paper we had an incorrect value for $\bar J_4$.  (The calculation of $\bar J_4$ was indeed tricky.)  We also incorrectly assumed that in equation (30) $r$ was equal to $s-1$.

\vfill\eject

\centerline{Appendix}

\ \ \ \ \ We consider the dimer problem on a one dimensional lattice with $N$ vertices.  These vertices may then be labeled by the integers $1, 2, ..., N$.  The generic located tile may be written $s_{ij}$ where $1 \le i < j \le N$, and clearly
\[	s_{ij} = \{ i,j \}  \eqno (A1)  \]
$f_0$, here, is exactly the constant function on tiles equal $1\big/(N-1)$.  $f$ is given as
\[  f(s_{ij}) =	\left\{ \begin{array}{cc} 
\frac 1 2 & {\rm if} \ \  j=i+1\\ 
 \\
\frac 1 2  & {\rm if} \ \ i=1 \ {\rm and} \ j = N \\
 \\
0  & otherwise  \end{array} . \right. \eqno (A2) \]
And by (10)
\[	v(s_{ij}) = f(s_{ij}) - f_0 \eqno(A3)	\]
We study $J_2 = 2N \bar J_2$.  We will actually be after $\bar J_2$ in the $N \ra \infty$ limit.

We refer to equation (28).  We find from [3] that $\psi'_c(s_1,s_2) = -1$ if $s_1 \cap s_2$ is non-empty, and zero otherwise.
\[	\bar J_2 = - \frac 1 2 \ \frac 1 N \ \sideset{}{'}\sum_{s_{ij},\, s_{i'j'}} \ \left( f(s_{ij}) - f_0\right) \left( f(s_{i'j'}) - f_0 \right) \eqno (A4) \]
where the prime on the summation sign is to represent the restriction to $s_{ij},\, s_{i'j'}$ with non-zero intersection.  We write $\bar J_2$ as the sum of six terms
\[	\bar J_2 = A + B + C + D + E + F  \eqno (A5)  \]
$A, B$, and $C$ are terms where $s_{ij}$ and $s_{i'j'}$ intersect in two vertices (i.e. $s_{ij} = s_{i'j'}$), and $D, E$, and $F$ are terms in which $s_{ij}$ and $s_{i'j'}$ intersect in a single vertex.  We expand the binomial in the sum of (A4), and $A$ and $D$ are terms quadratic in the $f$'s, $B$ and $E$, quadratic in the $f_0$'s, and $C$ and $F$ are crossterms involving one $f_0$, and one $f$.  We want the limits of these quantities as $N \ra \infty$, which we here collect.
\vfill\eject

\[ A \longrightarrow  - \frac 1 8 \eqno (A6)  \]
\[ B \longrightarrow  0 \eqno (A7) \]
\[ C \longrightarrow 0 \eqno (A8)  \]
\[ D \longrightarrow - \frac 1 4 \eqno (A9)  \]
\[ E \longrightarrow - \frac 1 2 \eqno (A10)  \]
\[ F \longrightarrow 1 \eqno (A11)  \]
From this will follow that $\bar J_2 \ra \frac 1 8$, the result in (38), for $d=1$.  We derive some of the limits in (A6) -(A11).

\noindent
\underline{Study of term B}

\[	B = - \frac 12 \ \frac 1 N \ \left( \frac{N(N-1)} 2 \right) \left( \frac 1{N-1} \right)^2 
\begin{array}[t]{c}
{\displaystyle\longrightarrow} \\
{\scriptstyle{N \ra \infty }} 
\end{array}
 \ 0  \eqno (A12)  \]
The last factor on the left side of the limit is $f^2_0$, and the next to last factor is the total number of possible located tiles.  We are here keeping terms in (A4) with $s_{ij} = s_{i'j'}$ arising from the product of the $f_0$'s.

\noindent
\underline{Study of term A}

\[	A = - \frac 12 \ \frac 1 N \cdot N \cdot \frac 1 4 = - \frac 1 8 \eqno(A13)   \]
The factor $\frac 1 4$ is $f^2$ on each dimer, and the factor $N$ is the total number of dimers.  We are here keeping terms in (A4) with $s_{ij} = s_{i'j'}$ arising from the product of the $f's$. $f$ is only non-zero on dimers.

\noindent
\underline{Study of term F}

\[	F = - \frac 12 \ \frac 1 N \cdot 2 \cdot N \cdot 2  \cdot (N-2) \cdot  \frac 1 2 \cdot \frac{-1}{N-1} \ra + 1. \eqno(A14) \]
The factor $- \frac 1{2N}$ comes directly from (A4); the next factor 2 is associated with the two cross terms, $ff_0$ and $f_0f$; the next $N$ arises because we pick $s_{ij} = s_{12}$ (we are looking at $f(s_{ij})f_0$, and using translation invariance of $f$, and the fact $f$ is only non-zero on dimers); 2 arises from the two choices $i'=1, \ i'=2$; $(N-2)$ is the number of choices for $j'$, once $i'$ is fixed, $\frac 1 2 \cdot \frac{-1}{N-1}$ is the non-zero value of $-f f_0$.

The other terms are similarly calculable.  This has been a typical computation for the $\bar J$'s.  There are lots of cases; and the number of cases grows for higher $\bar J_s$'s, as does the counting become trickier for each case, as $s$ grows.

\bigskip
\bigskip
\bigskip
\centerline{References}
\begin{itemize}
\item[[1]] Paul Federbush, Tilings With Very Elastic Tiles, Math--ph 0707.2525.
\item[[2]] Henryk Minc, An Asymptotic Solution of the Multidimensional Dimer Problem, Linear and Multilinear Algebra, 1980, {\bf 8}, 235-239.
\item[[3]] David C. Brydges, ``A Short Course in Cluster Expansions" in ``Phenomenes Critiques, Systems Aleatoires, Theories de Gauge, Part I, II" (Les Houches, 1984), 129-183, North Holland, Amsterdam, 1986.
\item[[4]] E.M. Fisher, Statistical Mechanics of Dimers on a Plane Lattice, Phys. Rev. 124 (1961), 1664-1672.
\item[[5]] P.W. Kasteleyn, The Statistics of Dimers on a Lattice, Physica 27 (1961), 1209-1225.
\item[[6]] S. Friedland, E. Krop, P.H. Lundow, K. Markstr\"{o}m, Validations of the Asymptotic Matching Conjectures, Math/0603001v2.

\end{itemize}

\end{document}